# Three-Dimensional Numerical Modeling of Shear Stimulation of Naturally Fractured Reservoirs

E. Ucar[1], I. Berre[1,2], and E. Keilegavlen[1]

[1] Department of Mathematics, University of Bergen, Bergen, Norway.

[2] Christian Michelsen Research, Bergen, Norway.

Corresponding author: Eren Ucar (eren.ucar@uib.no)

**Key Points:**

- New numerical framework for shear stimulation where fractures are represented as two-dimensional surfaces in three-dimensional rock matrix

- Fluid flow in fractures and matrix, fracture deformation, and rock matrix deformation are coupled

- It is shown how background permeability of the fractured rock highly affects permeability evolution and induced seismicity

**Abstract**

Shear-dilation based hydraulic stimulations enable exploitation of geothermal energy from reservoirs with inadequate initial permeability. While contributing to enhancing the reservoir's permeability, hydraulic stimulation processes may lead to undesired seismic activity. Here, we present a three-dimensional numerical model aiming to increase understanding of this mechanism and its consequences. The fractured reservoir is modeled as a network of explicitly represented large-scale fractures immersed in a permeable rock matrix. The numerical formulation is constructed by coupling three physical processes: fluid flow, fracture deformation, and rock matrix deformation. For flow simulations, the discrete fracture-matrix model is used, which allows the fluid transport from high-permeable conductive fractures to the rock matrix and vice versa. The mechanical behavior of the fractures is modeled using a hyperbolic model with reversible and irreversible deformations. Linear elasticity is assumed for the mechanical deformation and stress alteration of the rock matrix. Fractures are modeled as lower-dimensional surfaces embodied in the domain, subjected to specific governing equations for their deformation along the tangential and normal directions. Both the fluid flow and momentum balance equations are approximated by finite volume discretizations. The new numerical model is demonstrated considering a three-dimensional fractured formation with a network of 20 explicitly represented fractures. The effects of fluid exchange between fractures and rock matrix on the permeability evolution and the generated seismicity are examined for test cases resembling realistic reservoir conditions.

**1 Introduction**

An Enhanced Geothermal System (EGS) is created to utilize geothermal energy from subsurface regions where the reservoir rock is sufficiently hot but has an inadequate permeability to obtain commercial production rates. Shear-dilation based hydraulic stimulation (a.k.a. shear stimulation, low-pressure stimulation or hydroshearing) is an appealing method to create an EGS for reservoirs that have pre-existing fractures and are subjected to high deviatoric stresses (Murphy & Fehler, 1986; Pine & Batchelor, 1984). It relies on activation of naturally existing fractures through self-propping shear failure induced by fluid injections. The injection ensures to avoid the propagation of a single hydraulic fracture by keeping the pressure lower than the estimated minimum principal compressive stress. The shear failure leads to a mismatch of fracture asperities (i.e., shear dilation of the fractures) that results in permanent permeability enhancement in the reservoir, improving the reservoir connectivity and thus potentially allowing for economic production rates.

Although shear stimulation is successfully applied to several geothermal fields, vast amounts of induced seismicity have been reported during and after the stimulation process (Majer et al., 2007). Induced seismicity has generally been rarely felt by for EGS projects, and, hence, without economic consequences (Gaucher et al., 2015). However, some incidents with large seismicity such as the ones occurred in Basel and St.Gallen, lead to the cancellation of the projects (Edwards et al., 2015; Häring et al., 2008). In light of these incidents, there have been scientific efforts to better understand the relevant physical processes and mechanisms leading to induced seismicity (Baisch et al., 2010; Majer et al., 2007; McClure, 2015; Norbeck et al., 2016; Ucar et al., 2017). Also, mesoscale experiments have led to further understanding of the governing processes (Guglielmi et al., 2015).

Numerical modeling of the shear stimulation can be a powerful tool for estimating the potential performance of the stimulated reservoir and/or forecast possible undesired by-products of the stimulation process. However, due to the complex structure of the fractured rock and the number of coupled physical processes involved in the stimulation process, many modeling aspects remain uncertain. Challenges include the description of the mechanical and hydraulic behavior of the fractures considering that the activation of pre-existing fractures by fluid injection also leads to perturbation of the rock stress state, and capturing the resulting permeability evolution during shear stimulation. Thus, proper modeling requires coupling of three main hydro-mechanical processes: (1) fracture deformation, (2) fluid flow in fractures and within the porous rock matrix, and (3) alteration of stress state and deformation of the rock matrix in a three-dimensional (3-D) setting.

Accurate modeling of fracture deformation and permeability alteration is a crucial task as fractures are the main flow channels due to their greater permeability with respect to that of the rock matrix. Moreover, as demonstrated by several laboratory experiments (Bandis et al., 1981; Bandis et al., 1983; Barton et al., 1985; Goodman, 1976), fractures are more deformable than the rock matrix, and their alteration takes various forms, such as shear deformation, normal deformation, and dilation. While the linear elasticity assumption is generally sufficient for modeling the deformation of the rock matrix (Jaeger et al., 2007), experimental data show a nonlinear relation between stress and fracture deformation in the normal direction of the fracture (Goodman, 1976). This behavior is described with a hyperbolic formula by Bandis et al. (1981) and Barton et al. (1985). In the shear direction, fracture deformation is observed to have a linear relationship with applied shear stress before yielding, while a complex fracture deformation behavior has been detected after yielding (Barton et al., 1985). Shear displacement of the fracture surfaces leads to a dilation (a.k.a. shear dilation) in the normal direction as the fracture is forced to dilate when the rough fracture surfaces slide over each other. In hard rocks, the fracture is kept open, without the need for proppants upon the termination of injection, due to the strength of the contacting asperities. Apart from the classic Barton-Bandis joint deformation model for fracture deformation, its variant by Willis-Richards et al. (1996) has been widely used in the past decade (Kohl & Mégel, 2007; McClure & Horne, 2011; Rahman et al., 2002).

Although dynamic processes in an EGS setting are dominated by the fractures due to their higher permeability and deformation tendency, they have been implemented using simplified approaches for their physical and material complexity, with popular modeling choices generally falling into two classes. The first method is to define an equivalent continuum of both fractures and rock matrix where the effects of fractures on both the fluid flow and deformation are captured implicitly (Jeanne et al., 2014; Wassing et al., 2014). While more computationally efficient, the continuum models suffer from an important limitation due to insufficient representation of the fracture network where discrete effects of the fractures can be missed. A second common option is to model the fluid flow considering a discrete network of fractures immersed in an impermeable rock matrix (Baisch et al., 2010; Bruel, 2007; Kohl & Mégel, 2007; McClure & Horne, 2011; Willis-Richards et al., 1996). Although this is a common assumption in EGS applications, it has the major drawback of discarding that the rock will be permeable due to fine-scale fractures and pores in the formation surrounding the explicitly represented fractures. In practice, natural fractures occur at all scales (Berkowitz, 2002), and while the large-scale fractures dominate the fluid flow, the fine-scale fracture can have significant effects on the pressure distribution, thus on the stimulation. Leakage of fluid into the rock matrix can be represented by heuristic approaches, (Norbeck et al., 2016; Tao et al., 2011), but these studies are limited to two-dimensional (2-D) domains and simplified models for matrix flow and matrix-fracture interaction.

In this paper, we develop an approach to systematically handle flow and deformation of explicitly represented fractures, as well as the interaction with flow and deformation of the surrounding rock matrix. To explicitly model fractures at all scales is not computationally feasible, and also impractical due to a lack of data, in particular on the distribution of smaller scale fractures. A compromise is to apply a discrete fracture-matrix (DFM) approach, mainly developed to model flow in fractured rocks. In this framework, large-scale fractures constituting the main structural constraints on the processes, are represented explicitly, while surrounding regions (including fine-scale fractures) are represented by averaged quantities (Karimi-Fard et al., 2003; Sandve et al., 2012). When extended to also accommodate deformation of the rock matrix, the DFM model provides a natural framework for modeling the interplay between fluid pressure in the fracture and the rock-mechanical response. To include deformation of pre-existing fractures, fluid flow and deformation are coupled with the Barton-Bandis joint deformation model (Bandis et al., 1981; Bandis et al., 1983; Barton et al., 1985; Barton & Choubey, 1977). We present numerical simulations of stimulation of a complex 3-D fracture network that highlight the critical impact of matrix permeability on the effect of the stimulation process, including permeability enhancement and induced seismicity.

## 2 Governing Equations

The fractures of geothermal reservoirs in crystalline rocks are expected to have a first-order effect on the governing physical processes, and they are observed in various scales. To preserve the heterogeneities caused by the multi-scale fractures, a discrete fracture-matrix (DFM) representation of the fractured rock is applied here. In DFM models, the reservoir is treated as a combination of explicitly represented fractures and a surrounding rock matrix, which could implicitly incorporate the effects of fine-scale fractures. Conceptually, large-scale fractures are modeled as lower- dimensional objects, i.e., surfaces in the 3-D rock matrix, as illustrated in Figure 1.

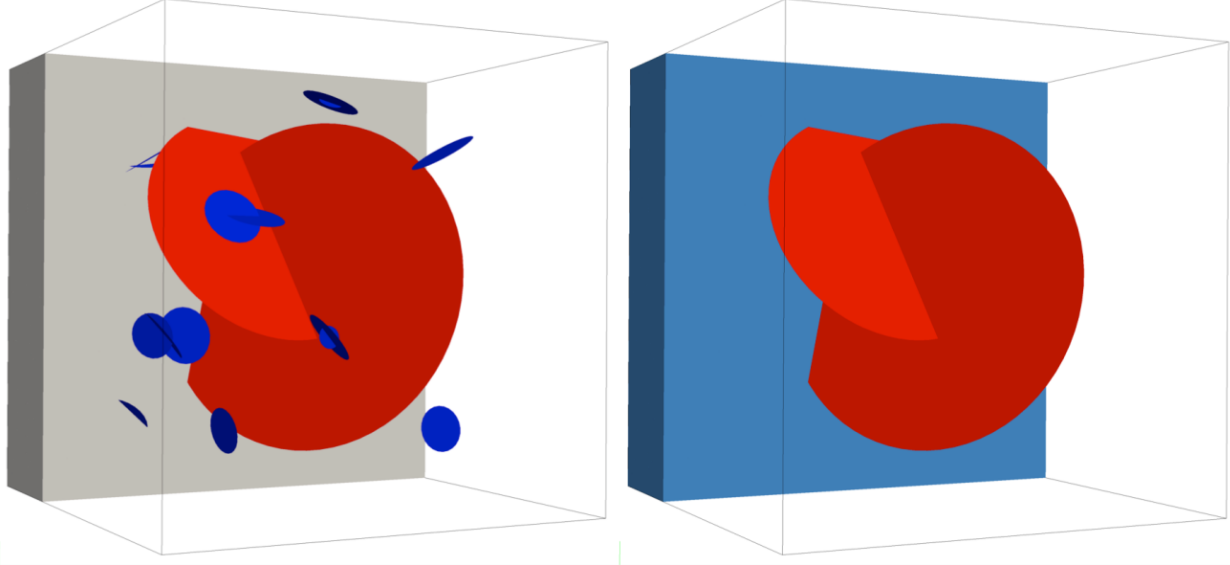

**Figure 1**. (Left) Demonstration of naturally fractured formations: the fractured rock consists of large-scale fractures (red), fine-scale fractures (dark blue) and the rock matrix surrounding the fractures (grey). (Right) Conceptual DFM rock model: the large-scale fractures (red) are explicitly represented while the fine-scale ones and the rock matrix are upscaled into continua (light blue)

The rock surrounding the explicitly modeled fractures, i.e., the rock matrix, will be denoted by $\Omega$. The rock matrix is modeled as a deformable porous media, with a permeability incorporating primary permeability due to the original porosity of the rock, as well as secondary permeability due to fine-scale fractures. The deformation of the matrix is modeled be linearly elastic, and, as we consider crystalline basement rock types, we assume that poro-elastic effects in the matrix are negligible. The fractures, denoted by $\Gamma$, are considered as void spaces created by two rough fractures surfaces that are in contact with each other. In this study, the main physical processes are assumed to be fracture deformation, fluid flow and rock matrix deformation. We do not include thermal effects in our modeling, i.e., the reservoir temperature is assumed to be constant during the stimulation. However, as the effect of the stimulation process will be illustrated by simulation of tentative reservoir production scenarios we will, for the sake of completeness, also present the heat transfer equation in the following.

2.1 Fracture failure and permeability alteration

The stress state of pre-existing fractures depends on the fracture orientation with respect to the anisotropic background stress conditions. According to the stress state of the fractures and fracture characteristics, we consider three types of fracture deformation: shear displacement of the fracture, normal dilation of the fracture due to shear displacement, and normal displacement of the fracture due to the elastic behavior of the rock.

Shear failure occurs when shear stress acting on a fracture is higher than the resistance of the fracture to slip. This resistance, $\tau_R$, can be estimated with the widely used Mohr-Coulomb criterion, which combines cohesion forces and frictional forces acting on a fracture surface as

$$\tau_R = S_0 + \mu \sigma_{n,eff}, \qquad (1)$$

where $S_0$ is the strength of the rock in absence of any normal stresses and $\mu$ is the friction coefficient. Defining the compressive stresses as positive, the effective normal loading, $\sigma_{n,eff}$, which creates frictional resistance in a pressurized fracture, can be calculated as the difference between the normal traction $\sigma_n$ and the pressure $p$. By injecting fluid into the fracture, the resistance $\tau_R$ is decreased, thus shear failures are facilitated.

For the sake of simplicity, we assume zero cohesion forces in this study. The friction coefficient is initially set equal to the static friction, $\mu_s$. After an element is exposed to a shear failure, its friction value is decreased to a dynamic friction value, $\mu_d$. This approach for friction modeling is known as a static/dynamic friction model and is typically denoted as an 'inherently discrete model' because the strength of the failing elements drops discontinuously with the slip (Rice, 1993). Although the model lacks convergence properties, it is reported to provide qualitatively acceptable results (McClure & Horne, 2011). More advanced constitutive laws, such as such as the rate- and state-dependent friction model for fracture behavior (Dieterich, 1979; Ruina, 1983), can also be integrated for more accurate modeling. For the fracture faces fulfilling the Mohr-Coulomb criterion, the shear displacement $\Delta d_s$ occurs in the direction tangent to the fracture surfaces. It can be approximated by using the excess shear stress concept (Rahman et al., 2002), which exploits linear elasticity, as

$$\Delta d_s = \frac{\Delta \tau}{K_s'}, \tag{2}$$

where $\Delta \tau$ is the excess shear stress and $K_s'$ is the shear stiffness per area of the fractures and is taken as a constant. The excess shear stress can be calculated as

$$\Delta \tau = \tau - \tau_R \tag{3}$$

where $\tau$ is the shear stress on the fracture surfaces. Further, the strength of the seismicity associated with the induced shear displacements is evaluated by calculating the seismic moment, $M_0$, as

$$M_0 = \int_A G \Delta d_s \, dA, \tag{4}$$

where $G$ is the shear modulus and $A$ is the slip area.

The second type of deformation is the elastic normal deformation due to the effective normal stress acting on the fracture surfaces. Bandis et al. (1983) and Barton et al. (1985) has suggested a model where it takes the form of

$$\Delta E_{n,rev} = \frac{\sigma_{n,eff}}{K_n' - \frac{\sigma_{n,eff}}{\Delta E_{max}}}, \tag{5}$$

where $\Delta E_{n,rev}$ is the reversible normal deformation, $K_n'$ is the normal stiffness per area and $\Delta E_{max}$ is the maximum possible closure. Although the Barton-Bandis joint deformation model defines $K_n'$ as a function of deformation and initial stiffness, for simplicity, we consider $K_n'$ to be constant and equal to the initial normal stiffness per area for each fracture in the following.

While the normal loading interacts continuously and elastically with the fracture aperture, shear slip results in a corresponding irreversible aperture change. The shear displacement changes the surface characteristics of the fracture irreversibly due to the asperity movement between fracture surfaces. Dilation occurs in a direction normal to the fracture surfaces, providing additional void space. The increase in the aperture caused by dilation is modeled by the following linear relation (Barton et al., 1985; Willis-Richards et al., 1996):

$$\Delta E_{n,irrev} = \Delta d_s \tan \varphi_{dil}, \tag{6}$$

where $\varphi_{dil}$ is the dilation angle, which in this study is assumed to be constant. Combining the above-mentioned deformations, the mechanical aperture, $E$, can therefore be written as

$$E = E_0 - \Delta E_{n,rev} + \Delta E_{n,irrev}, \tag{7}$$

where $E_0$ is mechanical aperture measured under zero stress conditions.

The aperture of real fractures varies in space and is affected by several parameters such as wall friction and tortuosity (Chen et al., 2000), which can be condensed into a single parameter called the joint roughness coefficient (JRC) (Barton & Choubey, 1977). The JRC range between 0 and 20 and can be measured experimentally or by comparison with existing JRC indices given by Barton and Choubey (1977). Here, we make a distinction between mechanical aperture, $E$, and hydraulic aperture, $e$, by considering the following relation suggested by Barton et al. (1985)

$$e = \frac{E^2}{JRC^{2.5}}. \tag{8}$$

Note that the units in equation (8) are in $\mu$m and the equation is valid only for $E \geq e$; thus, a maximum threshold, $E_{max}$, for mechanical aperture is enforced in the modeling. Finally, The permeability of the fractures, $K_f$, is associated with the hydraulic aperture of the fractures through the following 'cubic law' (Jaeger et al., 2007)

$$K_f = \frac{e^2}{12}. \tag{9}$$

2.2 Conservation of mass

We subdivide the fractures into two categories. The first type is the large-scale explicitly represented fractures (red fractures in Figure 1), which contribute dominantly to fluid flow due to their higher permeability values. The second type is the fine-scale fractures (blue fractures in Figure 1 (Left)). These fractures, although having relatively small size, contribute to the overall reservoir permeability. Therefore, in order to represent fine-scale fractures and pores, the rock matrix surrounding the large-scale fractures is modeled as a porous medium that has lower permeability.

The fluid flow in the matrix is governed by the mass conservation equation for an isothermal, single-phase, and slightly compressible fluid, which can be written as

$$\phi c_f \frac{\partial p}{\partial t} + \nabla \cdot \boldsymbol{w} = q, \tag{10}$$

where $\phi$ is the porosity of the rock, $c_f$ is the compressibility of the fluid, $t$ is time, $p$ is pressure, $\boldsymbol{w}$ is Darcy velocity, and $q$ is a source term. Neglecting gravitational effects, the Darcy velocity can be written as

$$\boldsymbol{w} = -\boldsymbol{K}\nabla p, \tag{11}$$

where the linear coefficient $K$ is the ratio between the intrinsic permeability of rock matrix and fluid viscosity. The matrix permeability should be computed according to the distribution of fine-scale fractures, see e.g., Lee et al. (2001).

For the fluid flow in the fractures, the mass conservation equation is customized to include the effect of pressure to the mechanical aperture that is explained in Section 2.1. Specifically, the porosity and permeability terms in equations (10) and (11) are modified according to the deformation of mechanical apertures. For the porosity of the fractures, the effect of mechanical aperture change is incorporated by using the mechanical aperture measured under zero stress condition, $E_0$. If $E_0$ represents the total volume of fractures, the porosity of fractures is calculated as $E/E_0$. The term, $K$, is considered isotropic, and calculated as the ratio between $K_f$ and fluid viscosity where the permeability of the fractures, $K_f$, is calculated as described in Section 2.1.

### 2.3 Conservation of momentum

Mechanical equilibrium of the rock matrix is governed by the balance of linear momentum equation under the quasi-static assumption; that is,

$$\nabla \cdot \boldsymbol{\sigma} = 0 \text{ on } \Omega, \tag{12}$$

where $\boldsymbol{\sigma}$ is the Cauchy stress tensor. Following Hooke's Law, we assume the components of the strain tensor are linearly related with the components of the stress tensor. Then, the stress tensor for the simple case of isotropic medium can be written as

$$\boldsymbol{\sigma} = 2G\boldsymbol{\varepsilon} + \lambda \, tr(\boldsymbol{\varepsilon}) \, \boldsymbol{I}, \tag{13}$$

where $G$ and $\lambda$ are the Lamé constants, $G$ being the shear modulus. Under the small-strain assumption, the linearized strains, $\boldsymbol{\varepsilon}$, are defined as the symmetric part of the displacement gradient

$$\boldsymbol{\varepsilon} = \frac{\left(\nabla \boldsymbol{u} + (\nabla \boldsymbol{u}^T)\right)}{2}, \tag{14}$$

where $\boldsymbol{u}$ is the displacement.

### 2.4 Mechanics of fractures

The fractures are modeled as two-sided co-dimension one inclusions (2-D surfaces) in the interior of the domain (3D rock matrix). In line with previous studies (Aagaard et al., 2013; Crouch & Starfield, 1982), the two surfaces of fracture are modeled as face pairs that have positive and negative sides. The face pairs are integrated as internal boundary conditions to the momentum balance equations, which are introduced in Section 2.3, by using the method developed by Ucar et al. (2016). The discontinuity relations due to the fracture deformation (equation (2), (5), and (6)) are defined as

$$\boldsymbol{u}_+ - \boldsymbol{u}_- = \Delta \boldsymbol{u}_f \text{ on } \Gamma \text{ where } \Delta \boldsymbol{u}_f = \boldsymbol{n}_+(\Delta E_{n,rev} + \Delta E_{n,irrev}) + \boldsymbol{\zeta}_+ \Delta d_s. \tag{15}$$

Here, $\boldsymbol{u}_+$ and $\boldsymbol{u}_-$ are the displacements on the positive and negative side of the fracture surfaces and $\Delta \boldsymbol{u}_f$ is the displacement jump vector, which is defined by the normal and shear displacement of the fracture. $\boldsymbol{n}_+$ denotes the unit vector defining the normal direction of the positive side of the fracture and $\boldsymbol{\zeta}_+$ denotes the unit vector defining the shear direction of the positive side of the fracture. In addition to jump conditions, the tractions on the fracture surfaces are continuous and satisfy equilibrium:

$$\boldsymbol{T}_+(\boldsymbol{n}_+) + \boldsymbol{T}_-(\boldsymbol{n}_-) = 0 \text{ on } \Gamma, \tag{16}$$

where $\boldsymbol{n}_-$ denotes the unit vector defining the normal direction of the negative side of the fracture. And the surface tractions are defined as

$$\boldsymbol{T} = \boldsymbol{\sigma} \cdot \boldsymbol{n}, \tag{17}$$

for any surface which has unit normal vector $\boldsymbol{n}$. We enforce the Kuhn-Tucker conditions of contact mechanics (Wriggers, 2006) as no penetration occurs between fracture faces and the effective normal traction stays compressive at the fracture surface.

### 2.5 Heat transfer

To illustrate the effect of the stimulation, we will consider a production scenario where water injected in a well, is transported through the reservoir and production of hot water is obtained from two production wells. Under the assumptions of incompressible fluid without any phase change, incompressible matrix, and local thermodynamic equilibrium between rock matrix and fluid, the heat transfer equation can be written as

$$\rho_{eff} c_{p,eff} \frac{\partial \theta}{\partial t} + \rho_f c_{p,f} \boldsymbol{w} \cdot \nabla \theta - \nabla \cdot (\boldsymbol{\kappa}_{eff} \nabla \theta) = f_{eff}, \tag{18}$$

where $\theta$ stands for both fluid and rock matrix temperature due to the local equilibrium assumption. Here, the effective heat capacity per volume, $\rho_{eff} c_{p,eff}$, effective thermal conductivity, $\boldsymbol{\kappa}_{eff}$, and total heat sources, $f_{eff}$, defined as

$$\begin{aligned} \rho_{eff} c_{p,eff} &= (1-\phi)\rho_r c_{p,r} + \phi \rho_f c_{p,f}, \\ \boldsymbol{\kappa}_{eff} &= (1-\phi)\boldsymbol{\kappa}_r + \phi \boldsymbol{\kappa}_f, \\ f_{eff} &= f_r + f_f, \end{aligned} \tag{19}$$

where $\rho_f$ is fluid density, $\rho_r$ is density of rock matrix, $c_{p,r}$ is heat capacity of rock matrix, $c_{p,f}$ is heat capacity of fluid, $\boldsymbol{\kappa}_r$ and $\boldsymbol{\kappa}_f$ are the thermal conductivities of rock and fluid, and $f_r$ and $f_f$ are heat source terms for rock matrix and fluid phase, respectively. The first term in the equation (19) is known as the storage term, the second term is the advective term, $\mathcal{H}_\mathcal{F}^{adv}$, and the third term is the diffusion term, $\mathcal{H}_\mathcal{F}^{diff}$.

## 3 Numerical Implementation

In this section, we present the numerical formulations and discretizations for the differential equations presented in the previous section. We first start with the introduction of the grid structure and continue with the space discretization for flow and linear elasticity equations. We also provide brief discretization details for heat transfer problem. Then, the section closes with the description of the numerical coupling approach and implementation notes.

### 3.1 Grid structure

The computational domain, $\Omega$, is divided into tetrahedral control volumes, $\Omega_i$, to represent the rock matrix by using Gmsh (Geuzaine & Remacle, 2009). The same computational grid is used for both the discretization of fluid flow and mechanics. The primary variables for both the problems, i.e., pressure ($p$) and deformation ($\boldsymbol{u}$), are defined at centroids of the control volumes. The grid structure for the rock matrix is created such that the faces of the tetrahedral volumes conforms to the fracture faces; that is, the fractures coincide with the faces of the computational cells. For the fracture discretization, we make slightly different modifications to the computational grid for the flow and the mechanics computations. For the flow problem, a hybrid approach is used to modify the grid, where the fractures are considered as lower dimensional objects (Karimi-Fard et al., 2003; Sandve et al., 2012). The fractures are converted into hybrid cells to represent fractures by assuming that the fracture is centered at the face and assigning to it an aperture; the hybrid fracture is shown in Figure 2a as a blue region. The primary variable for flow inside the fracture is associated with the hybrid cells centers (blue point in the blue region of Figure 2a). The volume of the hybrid cells is defined as the fracture volume, which is calculated by multiplying the area of the neighboring face by the aperture. For the mechanical problem, the fractures are two-sided co-dimension one inclusions in the interior of the domain. Thus, the computational grid is modified such that the fracture faces are duplicated to accommodate the displacement jump conditions, $\Delta \boldsymbol{u}_f$, as internal boundary conditions. Here, the primary variable, displacement, is defined at the face centers of the fracture faces as illustrated in Figure 2b. The details about the grid structure can be found in Ucar et al. (2016).

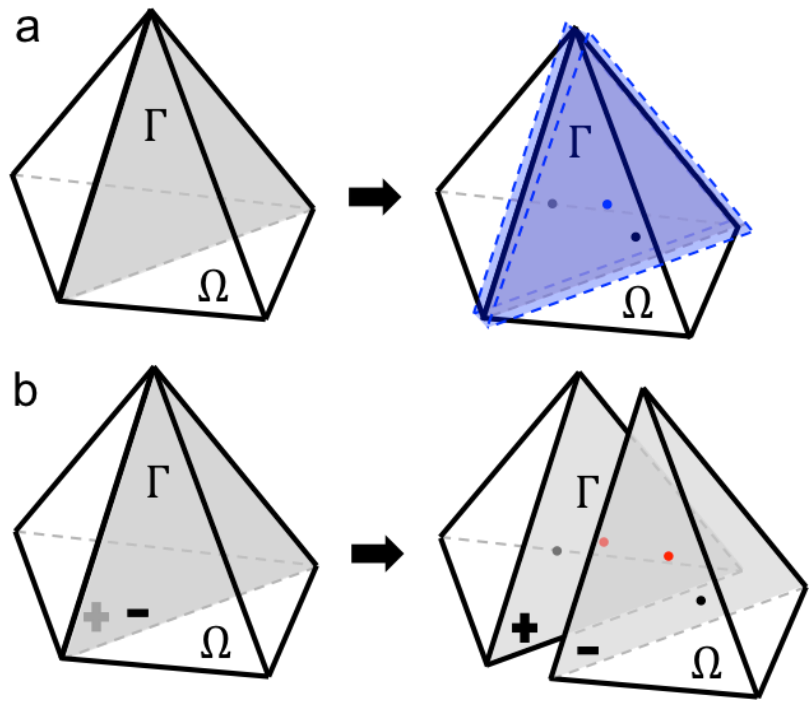

**Figure 2**. Grid structure and the modification of the grid for each problem. (a) The grid modification for flow problem. Hybrid cells, blue region, are created to represent fractures. The primary unknown, pressure, is defined at the hybrid cell center (blue dot). (b) The grid modification for mechanics problem. The faces that represent fractures are duplicated. The jump conditions, $\Delta \boldsymbol{u}_f$, are defined between face centers (red dots). For illustration purposes, we show a gap between two red dots, but there is no gap between the duplicated faces and the face centers in the computational mesh.

### 3.2 Spatial discretization

Finite-volume discretizations are used to approximate the solutions of both the fluid flow and mechanical deformation problem. By using cell-centered discretization for both of the problems, we can exploit the same data structure in computing both solutions. In this section, we present the discretization schemes.

#### 3.2.1 Discretization for the flow problem

The fluid flow equations in the DFM model are discretized using the two-point flux approximation (TPFA) presented in (Karimi-Fard et al., 2003). In TPFA, the discretization of the diffusive term in equation (10) starts with the integration over a cell volume, $\Omega_i$, and then applies the Gauss theorem to obtain surface integral as

$$\int_{\Omega_i} q \, dV = \int_{\Omega_i} -\nabla \cdot \boldsymbol{K} \nabla p \, dV = -\int_{\partial \Omega_i} \boldsymbol{n} \cdot \boldsymbol{K} \nabla p \, dS, \tag{20}$$

where $\boldsymbol{n}$ is the outward unit normal vector on boundary of cell volume, $\partial \Omega_i$. The flux across each face of a cell related to the edge $s$, $Q_s$, is expressed in terms of the pressure in the nearby cells, $n_c$,

$$Q_s = -\int_S \boldsymbol{n} \cdot \boldsymbol{K} \nabla p \, dS \approx \sum_{k=1}^{n_c} \gamma_k p_k, \tag{21}$$

where $\gamma_k$ is known as the face transmissibility. In the TPFA method, the flux between two neighboring control volumes $\Omega_i$ and $\Omega_j$ is approximated as

$$Q_{ij} \approx \gamma_{ij}(p_j - p_i), \tag{22}$$

where $p_i$ and $p_j$ are the pressures defined the centers of cell $i$ and cell $j$. The transmissibilities, $\gamma_{ij}$, corresponding to the face $\mathcal{F}$ between the two cells, depend only on adjacent cells of the edge, and are given by

$$\gamma_{ij} = \frac{\alpha_{i,\mathcal{F}} \alpha_{j,\mathcal{F}}}{\alpha_{i,\mathcal{F}} + \alpha_{j,\mathcal{F}}}, \tag{23}$$

where $\alpha_{i,\mathcal{F}}$ can be calculated as

$$\alpha_{i,\mathcal{F}} = \frac{A_{\mathcal{F}} \boldsymbol{n}_{\mathcal{F}} \cdot \boldsymbol{K}_i}{\boldsymbol{d}_{\mathcal{F}} \cdot \boldsymbol{d}_{\mathcal{F}}} \boldsymbol{d}_{\mathcal{F}}. \tag{24}$$

Here, where $A_{\mathcal{F}}$ is the area of the face, $\boldsymbol{n}_{\mathcal{F}}$ is the unit normal vector pointing outward from cell $i$, $\boldsymbol{K}_i$ is the permeability of cell $i$ and $\boldsymbol{d}_{\mathcal{F}}$ is the distance vector from the centroid of cell $i$ to the face centroid. Transmissibilities are calculated between neighbor rock matrix cells, between neighbor hybrid cells, and between neighbor rock matrix and hybrid cells.

The accumulation term of the mass conservation equation is approximated by the implicit backward Euler scheme.

#### 3.2.2 Discretization for mechanics problem

To approximate the solution of the mechanics problem, we use a cell-centered finite-volume method termed multi-point stress approximations (MPSA) (Nordbotten, 2014, 2015); specifically, we apply the weakly symmetric variant developed in Keilegavlen and Nordbotten (2017). The discretization is based on the integral form of the static momentum balance equation ignoring the inertia forces; that is,

$$\int_{\partial \Omega_i} \boldsymbol{T}(\boldsymbol{n}) \, dA = 0, \quad (25)$$

where $\boldsymbol{T}(\boldsymbol{n})$ are the surface traction vectors on the boundary of some domain, $\Omega_i$, with outward facing normal vector $\boldsymbol{n}$. In the discrete setting, the momentum conservation equation for each cell can be rewritten as

$$\int_{\partial \Omega_i} \boldsymbol{T}(\boldsymbol{n}) dA = \sum_j \int_{\mathcal{F}} \boldsymbol{T}(\boldsymbol{n}) dA = \sum_j \boldsymbol{T}_{i,j}, \quad (26)$$

where $\mathcal{F}$ is the shared boundary between adjacent cells of $i$ and $j$ and $\boldsymbol{T}_{i,j}$ is the surface stresses over the boundary $\mathcal{F}$. The discrete stress, $\boldsymbol{T}_{i,j}$, is defined as a linear function of displacements:

$$\boldsymbol{T}_{i,j} = \sum_k \tilde{t}_{i,j,k} \boldsymbol{u}_k, \quad (27)$$

where $\tilde{t}_{i,j,k}$ are referred as stress weight tensors and $\boldsymbol{u}_k$ are the displacements located at cell centers. The stress weight tensors are calculated by imposing continuity of surface stresses and displacements, where each component of displacement is approximated by a multilinear function of the spatial coordinates.

In a hydraulic stimulation setting, we assume that the rock matrix is at equilibrium initially. Any perturbation in the system is caused by the deformation of the fracture surfaces due to the pressurization. As we have presented in the previous section, the fracture deformation is in the form of displacement jumps between the fracture surfaces. MPSA provides the inclusion of jump conditions as internal boundary conditions, provided that an appropriate mesh is supplied (Figure 2b). During the discretization, the jump conditions denoted as $\Delta \boldsymbol{u}_f$, are defined as internal boundary conditions for the system and corresponding stress alterations are approximated. More detail on this coupling scheme can be found in Ucar et al. (2016).

### 3.2.3 Discretization for heat transfer problem

The heat transfer equation requires temporal discretization, the discretization for the diffusion term, $\mathcal{H}_{\mathcal{F}}^{diff}$, and the discretization for the advective term, $\mathcal{H}_{\mathcal{F}}^{adv}$. The temporal discretization is handled by the implicit backward Euler scheme. $\mathcal{H}_{\mathcal{F}}^{diff}$ has the same form as the diffusion term in the flow equation; therefore, it is approximated in the same manner by using TPFA, resulting in

$$\mathcal{H}_{\mathcal{F}}^{diff} \approx D_{ij}(\theta_i - \theta_j), \quad (28)$$

where the heat diffusivities between cells, $D_{i,j}$, are computed as in equation (24) with the effective thermal conductivity values replacing permeability values. The advective term, $\mathcal{H}_{\mathcal{F}}^{adv}$, that is the energy that is transported by the flow, is calculated with an upwind discretization. The amount of advective heat transfer over the face $\mathcal{F}$ between the two cells with index $i$ and $j$ is determined by the flux direction, $\boldsymbol{w}_{\mathcal{F}}$, by

$$\mathcal{H}_{\mathcal{F}}^{adv} \approx \boldsymbol{w}_{\mathcal{F}} \theta_{upwind,ij}, \quad (29)$$

where $\theta_{upwind,ij}$ is calculated as

$$\theta_{upwind,ij} = \begin{cases} \theta_i \text{ if } \boldsymbol{w}_{\mathcal{F}} \cdot \boldsymbol{n}_{\mathcal{F}} > 0, \\ \theta_j \text{ if } \boldsymbol{w}_{\mathcal{F}} \cdot \boldsymbol{n}_{\mathcal{F}} < 0. \end{cases} \quad (30)$$

While $w_\mathcal{F} \cdot n_\mathcal{F} > 0$ indicates that the flow direction is from cell $i$ to $j$, $w_\mathcal{F} \cdot n_\mathcal{F} < 0$ indicates that the flow occurs from cell $j$ to $i$. Details on the discretization for the heat transport can be found in Stefansson (2016).

### 3.4 Multi-physics coupling

The coupling of fluid flow, stress alteration and fracture deformation to the simulation of stimulation requires a two-stage algorithm for each time step, see flowchart in Figure 3. The first stage involves the propagation in time of the pressures via equation (10), and the computation of a corresponding new equilibrium state of the reversible aperture change, $\Delta E_{n,rev}$. The equilibrium is found by balancing fluid pressure at the fracture walls with the rock mechanical response to an aperture change, with the latter computed by (12)-(14) with boundary conditions taken as aperture change. This is a non-linear system, due to the fracture permeability being dependent on aperture, which we solve iteratively by sequentially coupled flow and rock mechanics solves. The states from the previous time step as initial guesses, while the iterations are terminated when the aperture change is smaller than a certain threshold.

If the first stage produces stresses on fracture faces that exceed the Mohr Column threshold, the second part of the algorithm is invoked to calculate the irreversible deformation of the fracture and the corresponding aperture change. For each face that violates the Mohr-Coulomb criterion, the shear displacements and corresponding irreversible normal deformations (dilations) on the fracture surfaces are calculated. Then the response from the rock matrix to this fracture deformation is again calculated from (12)-(14), with the fracture deformation as an internal boundary condition. The stress alterations in the domain can modify the shear and normal stresses at some locations; thus, further failures can be initiated by so-called 'slip avalanches'(Baisch et al., 2010). Therefore, after the slip of an element, the mechanical state of the system is recalculated and the Mohr-Coulomb criterion successively checked for all fractures. Mohr-Coulomb criterion check continues until additional displacements are no longer induced. When an equilibrium state is reached, the next time step is executed. In this step, the pressure relaxation associated with the dilation of fractures is not accounted. The fluid pressure is fixed throughout this second stage, corresponding to an assumption that irreversible shearing is instantaneous relative to the time scale of fluid flow.

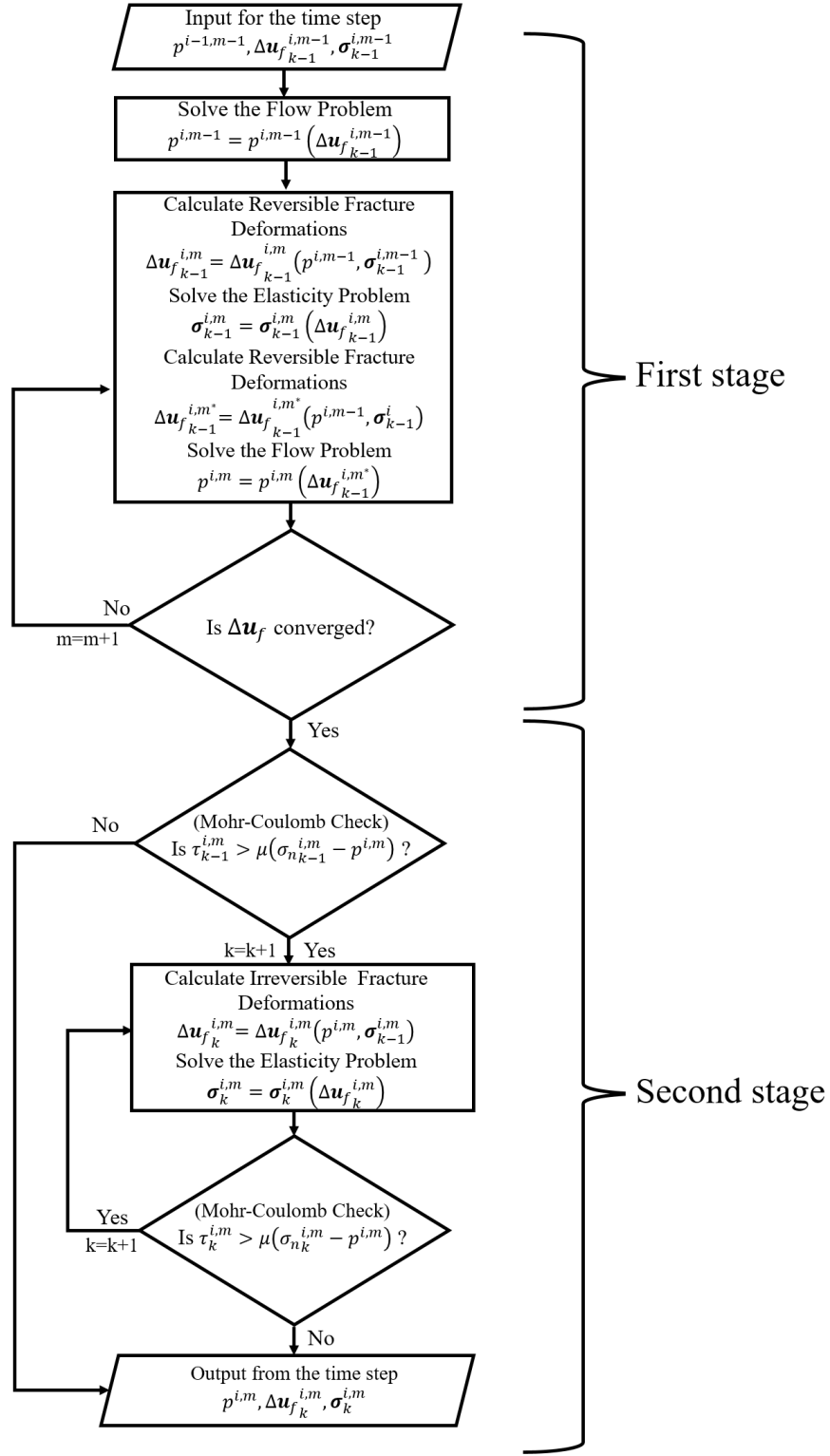

**Figure 3.** The two-stage solution procedure to model shear stimulation. The two-stage is applied for each time step.

### 3.5 Implementation

Our numerical model for shear stimulation is developed utilizing the MATLAB Reservoir Simulation Toolbox (MRST) (Lie, 2016; Lie et al., 2012). MRST is a free open-source reservoir simulator, which offers a wide range of data structures and computational methods. The software includes several add-on modules that one can use or couple with other modules for custom-made modeling. We couple two modules of MRST to implement our scheme. For flow simulations, the Discrete Fracture-Matrix module is used. The module provides control volume discretizations (two- and multi-point flux approximations) for fluid flow in fractured formations. Thanks to the representation of fractures as lower dimensional objects in the computational grids, the module is capable of modeling the interacting fluid flow in fractures and surrounding rock matrix. For the solution of linear momentum balance equation, the MPSA module is used. This module provides again cell-centered discretization of the governing equations, which facilitates the coupling between mechanical deformation and fluid flow. With this as a basis, fracture deformation is implemented, updating the discretization matrix to incorporate the internal boundary conditions for the fractures. Discretization of both mass and momentum balance equations lead to linear systems to be solved for pressure and displacement, respectively. For the simulations reported here, the flow system is solved with a direct solver, as implemented by the MATLAB backslash operator. The elasticity system is considerably larger, and we, therefore, solve it by an iterative approach, specifically using GMRES (Saad & Schultz, 1986) preconditioned with ILU.

## 4 Numerical Results

We show the applicability of our model through two simulations of stimulation of a 3-D reservoir hosting 20 explicitly represented fractures subjected to potential slip. The synthetic examples are constructed to resemble realistic scenarios and highlight the capabilities of the methodology to (1) couple fracture and matrix deformation in complex fracture networks; (2) capture the effects of fluid flow in both fractures and matrix. Our first case includes the shear stimulation of a fracture network in which the fractures have different orientations with respect to the principal stresses. Hence, the effects of fracture orientation on the shear stimulation can be observed. For further analysis of the effect of shear stimulation, we also apply the resulting reservoir structure to the analysis of a production scenario. The temperature profiles of the stimulated and non-stimulated reservoir after one year of production are presented. Moreover, we examine another scenario that has the same fracture network but different rock matrix permeability than the first example. This example shows the significance of mass exchange between fractures and the rock matrix to the stimulation results.

### 4.1 Problem construction

A synthetic reservoir with a complex network of 20 preexisting fractures is considered. The fractures are defined as 2-D circular (penny-shaped) planes of weaknesses in a 3-D elastic material that represents the rock matrix. The diameters of the fractures are chosen to vary between 1000 m and 3000 m. Dirichlet boundary conditions are defined for both the fluid flow and mechanical problem as constant pressure and zero displacement conditions, respectively. Notably, the computational domain is chosen large enough to minimize the spurious effects of boundary conditions. The initial conditions for both conservation equations are set to be constant pressure and zero displacement fields following the assumption of flow and mechanical equilibrium. In addition, following the most common conditions in EGS in Europe (Gaucher et al., 2015), the background stress regime is considered as strike-slip $S_H > S_V > S_h$, which depends on depth as

$$S_H = 90 + 0.002z,$$
$$S_h = 50 + 0.002z, \quad (31)$$
$$S_V = 70 + 0.002z,$$

where $z$ is the depth in meters and the units of $S_H, S_V$ and $S_h$ are in MPa. The fractures are oriented variously with respect to the background stress that creates distinct stress conditions for each fracture. The fracture orientation properties are listed in Table S1 in supporting information.

A vertical injection well, which penetrates five fractures, is located approximately in the middle of the domain. For simplicity, we assume a constant 40 MPa hydrostatic pressure in the reservoir and the injection is controlled by the rate. The injection rate is tuned such that the reservoir fluid pressure never exceeds the minimum principal stress, ensuring that tensile fractures would not propagate. Fluid is injected for a period of 24 hours. The injection is started with 4 kg/s and increased 1 kg/s every hour until 15 kg/s and kept constant at 15 kg/s.

The static and dynamic friction coefficients are assumed to be 0.6 and 0.55, respectively. For two of the fractures, higher friction values ($\mu_s$= 0.75 and $\mu_d$ =0.65) are used to ensure the system is steady state at the start of the fluid injection. The mechanical apertures under zero stress conditions are assumed to be constant and the same for all fractures. The initial permeability is calculated by using $E_0$ and the initial stress conditions. Since each fracture has different stress conditions due to their orientation (strike, dip angle and the depth), the initial permeability values are different for each fracture. The fracture network, the injection well and the initial permeability map can be seen in Figure 4. The physical properties of the fluid and fractured rock are motivated by the realistic measurements and are listed in Table S2 in the supporting information.

For the numerical discretizations, the time step and the grid structure have to be defined. The time step is chosen to be 1 hour. A single, unstructured computational grid is used for both the flow and the mechanics problem. The grid consists of 25592 unstructured tetrahedrons, which conforms the fracture faces. The number of triangles (faces of conforming tetrahedrons) that discretized the fractures is 6404. The grid structure is presented in Figure 5. And the grid structure is provided as data set in the supporting information.

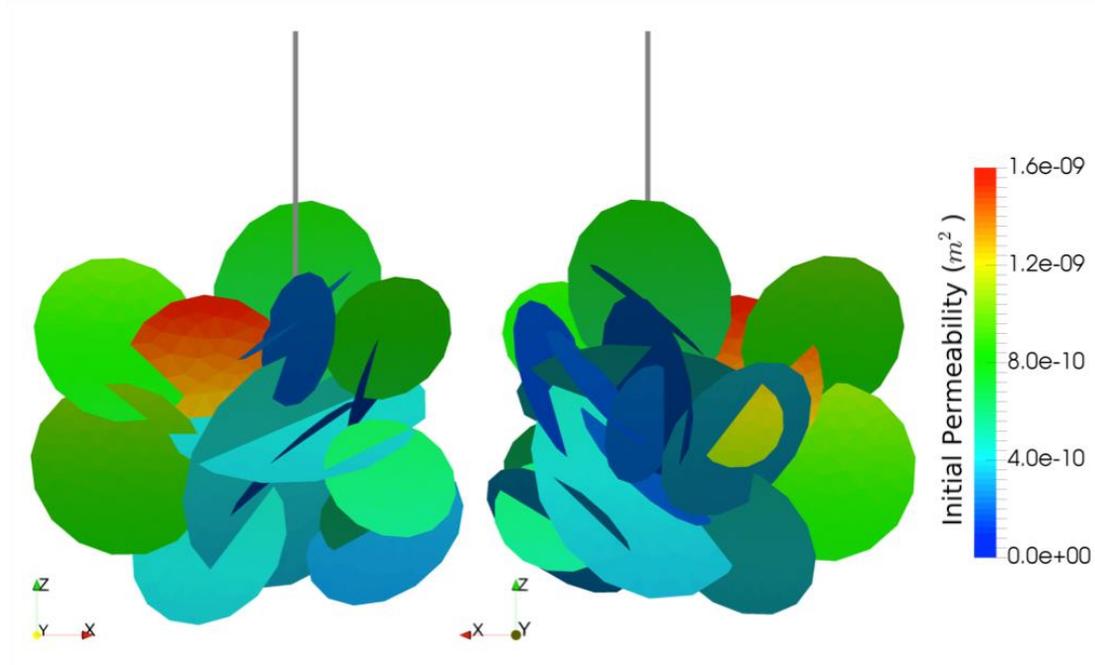

**Figure 4.** The fracture network, the injection well and the initial permeability map of the considered synthetic reservoir seen from two different angles.

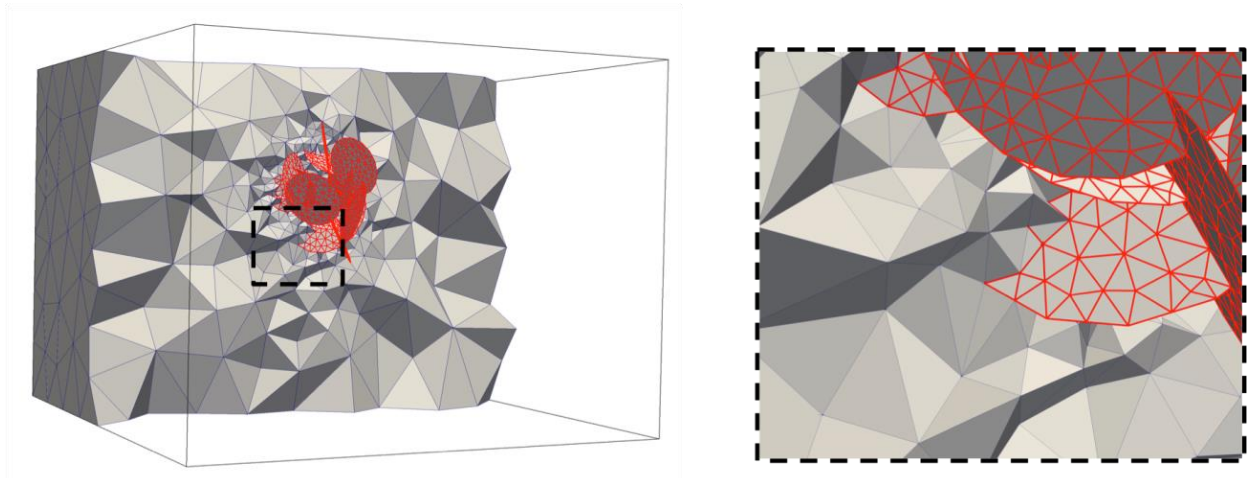

**Figure 5.** The details of the grid used for both the flow and the mechanics problem. (Left) The domain is cut approximately from the middle to show the grid structure both in the rock matrix and on the fractures. (Right) A closer look at the dashed area in the grid. The rock matrix is discretized by using unstructured tetrahedrons, and the fractures are discretized by triangles conforming tetrahedrons. The triangles discretizing the fractures are shown by red.

### 4.2 Numerical results

The section starts with the presentation of the stimulation results after one-day of injection for the synthetic reservoir presented in Section 4.1. We denote this numerical experiment as Case 1. For this case, we conduct further analysis on the effect of stimulation by demonstrating a one-year production scenario considering two different initial states corresponding to a non-stimulated and stimulated reservoir. We continue our numerical analysis with Section 4.2.2. This section includes the examination of the same setup with slightly different the rock permeability, which we denote as Case 2.

### 4.2.1 Permeability enhancement (Case 1)

As a beginning, the presented problem is examined with setting the matrix permeability to 2e-19 $m^2$. Figure 6 shows the results of the presented problem. After the stimulation process, all the fractures are exposed to a pressure change. The initial permeability combined with the orientations (the angle with respect to principle stresses) and the location (the depth) of the fractures determines the pressure distribution map and the generated seismic moments.

Seismicity is detected for 13 fractures out of 20 in this stimulation scenario. The maximum generated seismic moment rate is calculated as approximately 7.3e+9 Nm/s, and the minimum generated seismic moment rate is calculated as around 1e+4 Nm/s. We observed that high seismicity is generated at the beginning of the stimulation process and the seismicity decreases towards to the end of the stimulation. There are two reasons for this behavior: the first one is the fix of the injection rate at a constant value after the 12th hour and the second one is the leakage to the rock matrix. It is interesting to note that, if the orientation of a fracture results in high normal loading and/or low shear loading on the fracture, the corresponding initial permeability can be lower than that of the other fractures while still providing important pathways for fluid flow. This can be an essential aspect in terms of seismicity observations as fractures with a high normal load may not be visible in the seismicity recordings, but still can enhance the connectivity of the reservoir substantially.

The bottom part of Figure 6 shows the permeability enhancement map after the stimulation process. High permeability enhancement is observed at the fractures located at relatively low depth. The main reason is the less normal loading in the shallower regions according to the introduced stress field (equation (31)). The figure shows that initially, low permeable fractures are exposed to low permeability enhancement in general. However, the results also show that higher initial permeability of a fracture does not specifically lead to high permeability enhancement, as the primary mechanism for permeability enhancement is the shear dilation, which is strongly dependent on the orientation of the fracture. The average fracture permeability enhancement is also calculated for each fracture, and a one to 14-fold average permeability enhancement is observed over the one-day stimulation experiment.

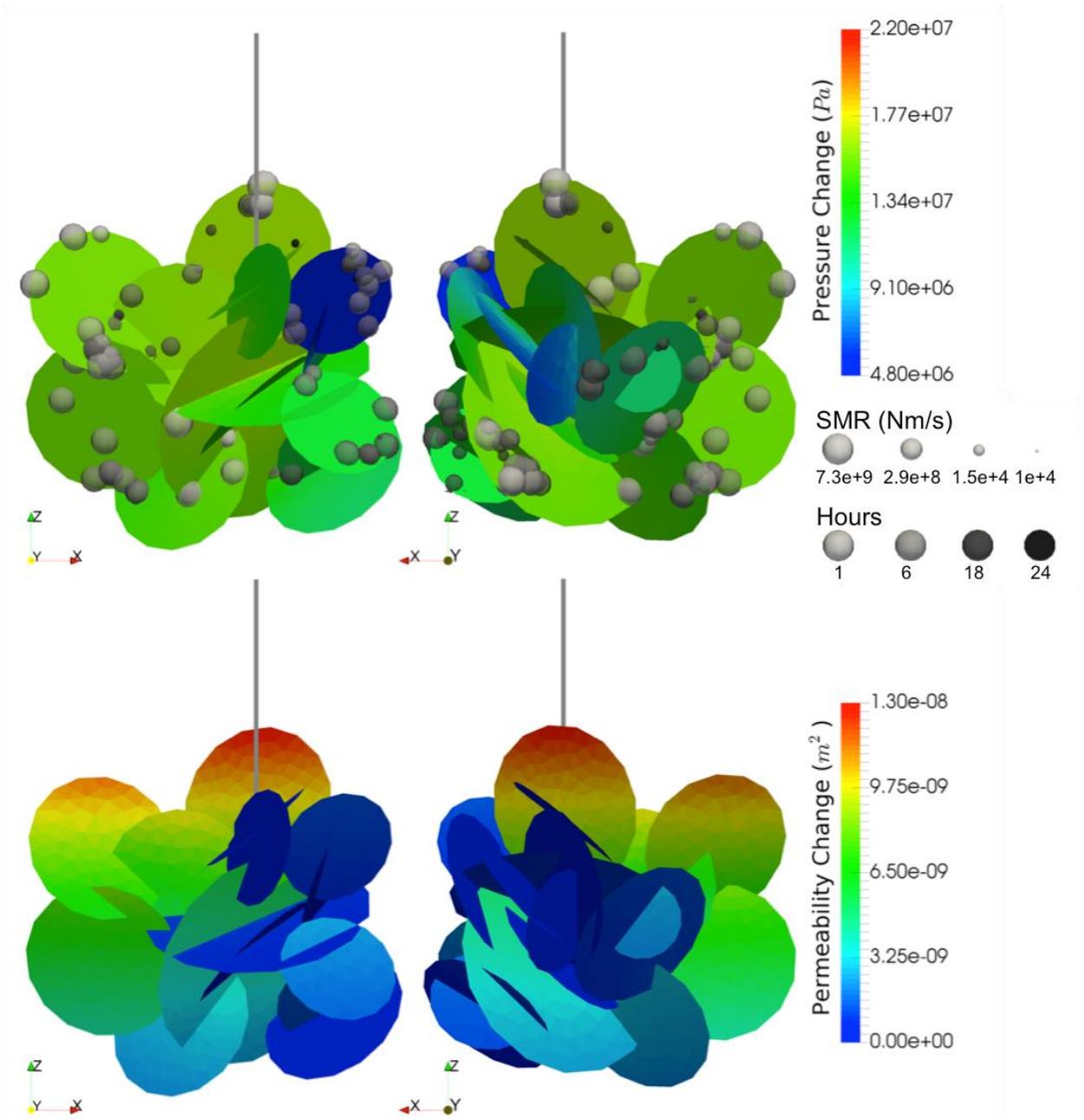

**Figure 6.** The results of the presented stimulation problem when matrix permeability is 2e-19 m$^2$. (Upper) The pressure change and the seismicity in the reservoir after one-day of stimulation shown in two views. The seismic moment rate (SMR) is shown with the spheres located at the faces where the largest displacement occurred for each fracture. The diameters of the spheres are associated with the size of SMR, while their colors are associated with the time step that SMR generated. The spheres are created with transparent color such that one can observe several seismic events located at the same point. (Bottom) The permeability change in the reservoir after one-day of stimulation shown in two views.

The main motivation for simulating the hydraulic stimulation process is to assess its impact on the reservoir performance in an energy production scenario. The DFM model applied herein can easily be applied in heat transport simulations using both initial and updated fracture apertures. To illustrate this, we show the effect of hydraulic stimulation on the temperature profiles for a one-year production scenario. In addition to the existing set-up, we introduce two production wells in the reservoir for the heat transfer analysis. In this example, the injection and the production wells are controlled with bottom-hole pressure as 50 MPa and 0.5 MPa, respectively. The boundary conditions for the heat transfer problem are constant temperature, which is taken equal to the initial reservoir temperature. The same domain and grid introduced in Section 3.1 are applied. The physical properties of the heat transfer problem are provided in Table S2 in the supporting information.

The upper part of Figure 7 shows the temperature profile of the fluid inside the fractures when we consider the non-stimulated state of the reservoir and the lower part of the figure belongs to same production scenario with the stimulated state of the reservoir is considered. Although included in the simulation model, the temperature distribution in the matrix is not shown for clarity of visualization. As expected, the hydraulic stimulation results in a wider sweep of the low-temperature front, thus enlarging the region from which high-temperature fluid is drawn towards the production wells.

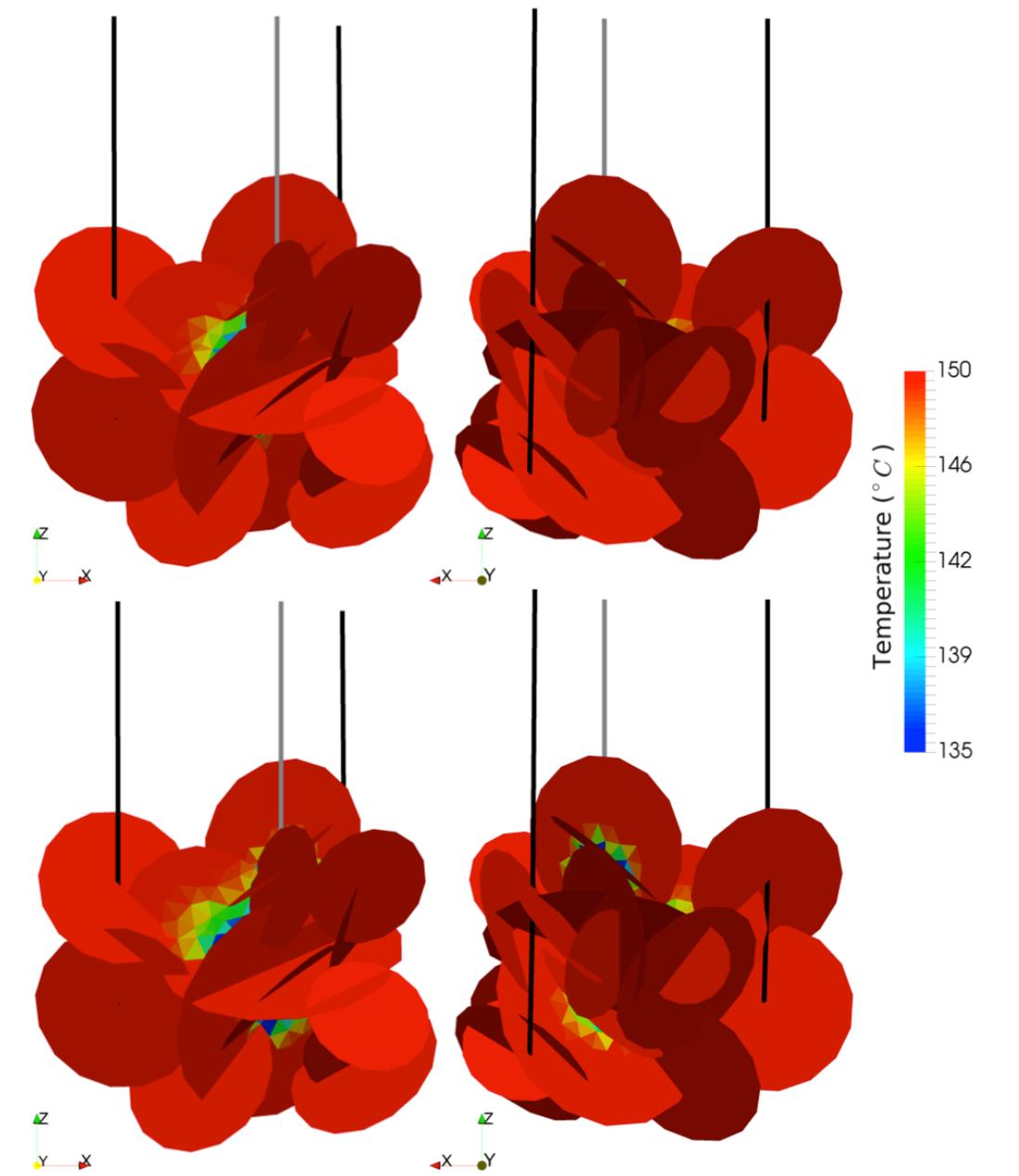

**Figure 7.** The temperature profiles after one year of production scenario for (upper) a non-stimulated and (bottom) stimulated reservoir. Black lines represent the production wells and the gray line the injection well.

4.2.2 Effect of matrix permeability on the permeability enhancement of fractures (Case 2)

Another numerical experiment is conducted for the same problem but with increasing the matrix permeability to 4e-19 $m^2$ to show both the capabilities of the model and the effect of the matrix permeability on the results. Figure 8 presents the pressure and permeability change for this problem. The change in pressure is observed to be less than to Case 1 due to the higher leakage to the rock matrix. Moreover, the number of fractures for which seismicity is observed is only six in this stimulation scenario while it is 13 for Case 1. The maximum generated seismic moment rate is calculated as approximately 3.9e+9 Nm/s, and the minimum generated seismic moment rate is calculated as around 3.4e+5 Nm/s. Although the behavior of permeability enhancement is similar to that of Case 1, the total permeability enhancement is observed to be lower, as expected. More specifically, the average permeability enhancement of the fractures is between one- and sevenfold.

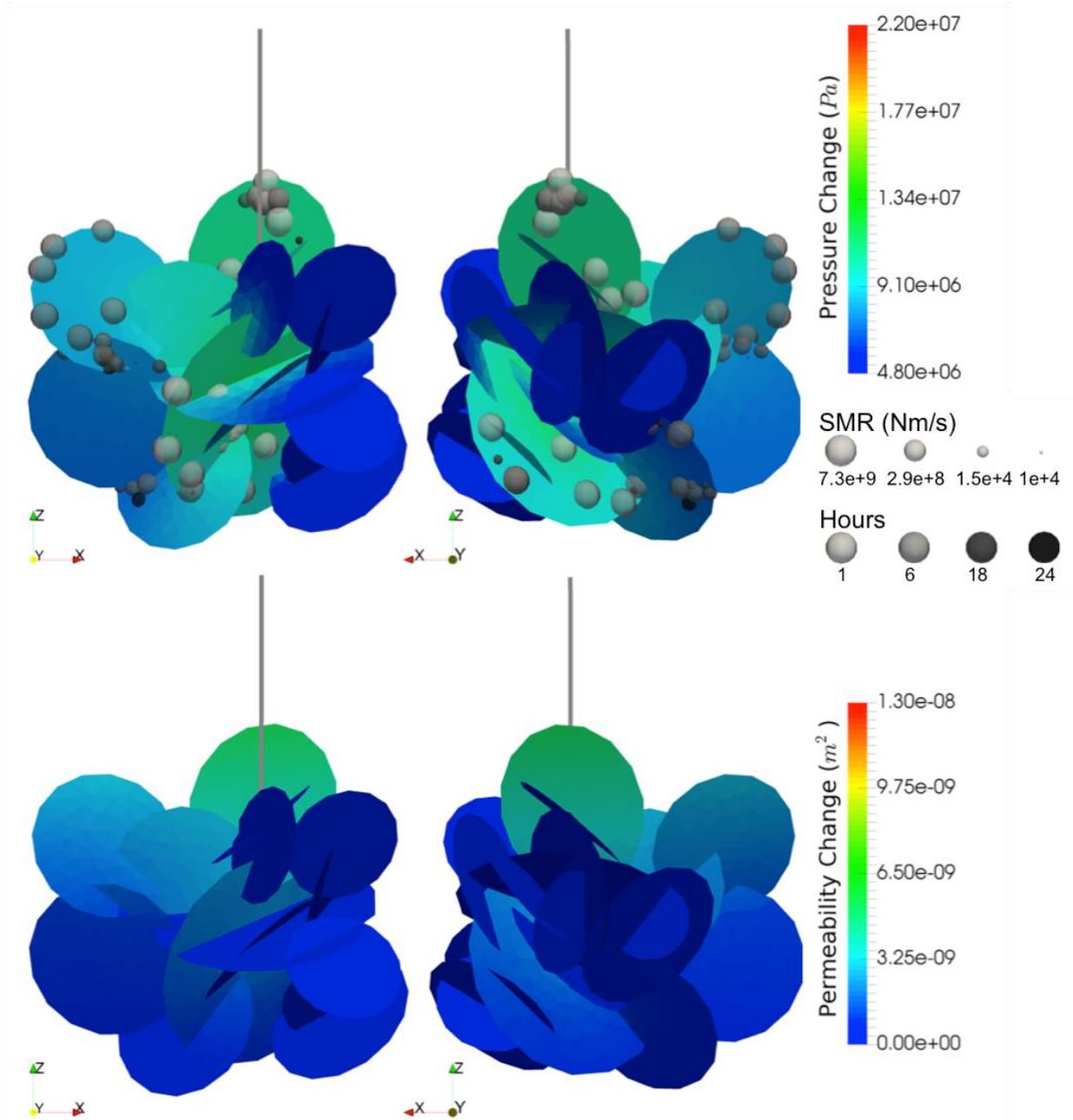

**Figure 8.** The results of the presented problem when matrix permeability is 4e-19 m2. (Upper) The pressure change and the seismicity in the reservoir after one-day of stimulation shown in two views. The seismic moment rate (SMR) is shown with the spheres located at the faces that the largest displacement occurred for each fracture. The diameters of the spheres associated with the size of SMR and the colors of the spheres are associated with the time step that SMR

generated. The spheres are created with transparent color such that one can observe several seismic events located at the same point. (Bottom) The permeability change in the reservoir after one-day of stimulation shown in two views.

To better investigate the difference between Case 1 and Case 2, we study the seismic moment rate for each fracture in a time step. Figure 9 presents the seismic moment rate for each fracture at each time step for both cases. Both the number of fractures that exposed seismicity and the amount of seismic moment rate is significantly lower for Case 2 than Case 1. The cumulative generated seismic moment is calculated as 5.25e+14 Nm for Case 1 and 2.23e+14 Nm for Case 2. A similar behavior of seismic moment rate is observed for both of the cases, as the seismicity is around the same range in the first 12 hours even though the injection rate is linearly increased in this period. The main reason of this is the increase in permeability in each time step. Although the amount of injected fluid is increasing, the available void space to host injected fluid is also increasing. The same reason is valid for the cause of seismic moment rate decrease the last 12 hours. In this period, the same amount is injected each time step, but due to the increased permeability, the pressure change is decreasing each time step. For both cases, seismicity is not observed after the termination of the injection well due to the high leakage to the rock matrix. In this case, the matrix permeability allows for sufficient fluid flow into the matrix to inhibit the adequate pressure build-up inside the fracture to cause after shut-in seismic events.

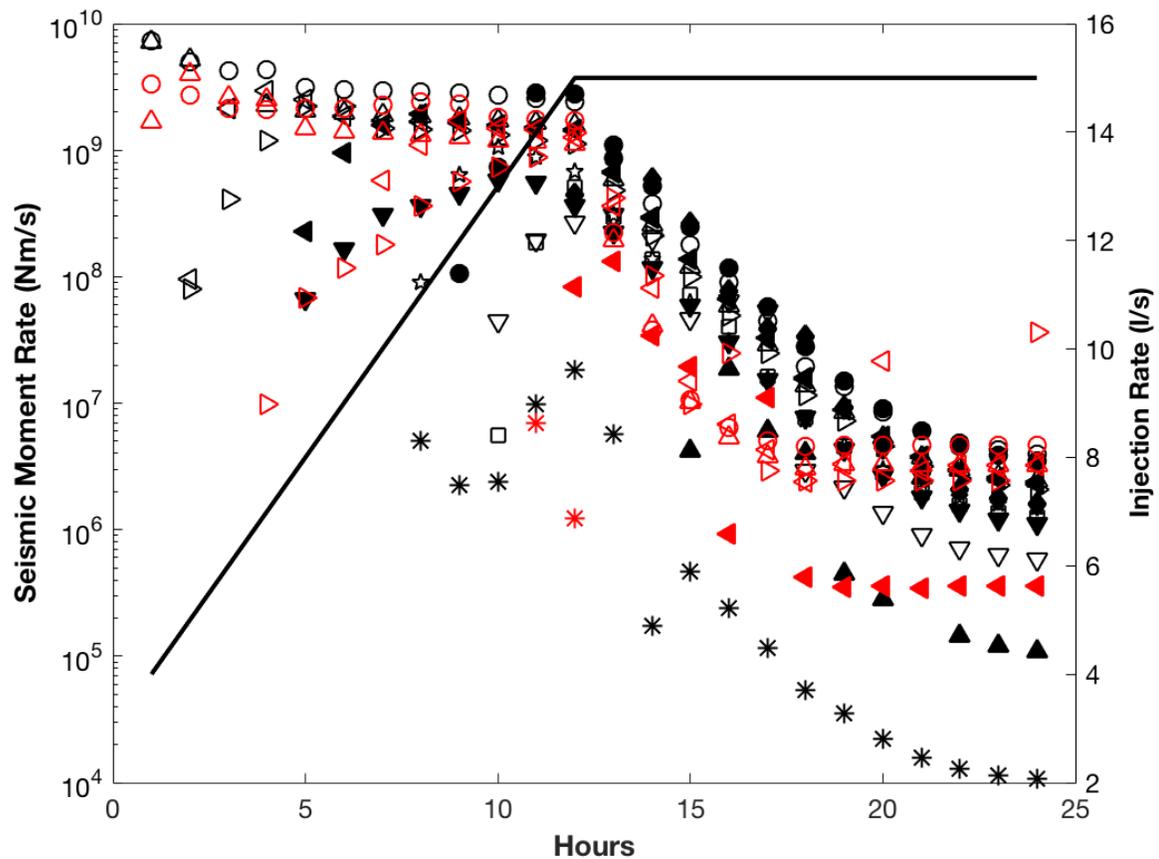

**Figure 9.** Moment rate of the induced seismicity and injection rate as a function of time. Each marker type represents a fracture. Black markers show the modeled seismicity for Case 1 and red markers show results for Case 2. The solid black line indicates the injection rate history.

## 5 Conclusions

A new computational model is presented to simulate shear dilation based hydraulic stimulations, aiming at assisting the assessment of geothermal reservoirs by forecasting the outcomes of a stimulation process, in terms of both permeability enhancement and induced seismicity.

The approach is based on a DFM model that subdivides the domain into the explicitly represented fractures and the rock matrix. Three main mechanisms are coupled in the model: fluid flow in fractures and matrix, rock matrix deformation, and fracture deformation, where the latter includes both slip and dilation as well as normal deformation of the fracture due to mechanical loading.

The developed model is investigated by conducting two simulations for a 3-D synthetic reservoir that hosts 20 fractures. The numerical experiments emphasize the effects of the complex structure of fracture network and the leakage to rock matrix to the permeability enhancement and induced seismicity. The impact of shear stimulation on the productivity of the reservoir is also demonstrated.

We believe that the capabilities of the presented methodology will be valuable in understanding governing mechanisms in stimulation of fractured geothermal reservoir. It has already been applied in investigating the effects of the normal closure of fractures on post-injection seismicity (Ucar et al., 2017). Our model can further be used for geothermal reservoir assessment by, for example, simulations of different injection scenarios (e.g., monotonic, cyclic, rate-controlled, pressure-controlled injection) for reservoirs structurally dominated by complex fracture networks that have various hydromechanical properties under different anisotropic stress conditions.

## Acknowledgments

Data associated with this study are available in the supporting information. The authors would like to thank Ivar Stefansson for the helpful suggestions for Section 3.2.3. The work was funded by the Research Council of Norway through grant no. 228832/E20 and Statoil ASA through the Akademia agreement.